\documentclass[aip]{revtex4-1}

\draft 

\usepackage{graphicx}
\usepackage{textcomp}

\begin{document}

\title[Modeling of O--X conversion in Pegasus]{Full-wave modeling of the O--X mode conversion in the Pegasus Toroidal Experiment}

\author{A. K\"ohn}
  \email{koehn@ipf.uni-stuttgart.de}
  \affiliation{Institut f\"ur Plasmaforschung, Universit\"at Stuttgart, Germany}
\author{J. Jacquot}
  \affiliation{IRFM, CEA, F-13108 Saint-Paul-lez-Durance, France}
\author{M.~W. Bongard}
  \affiliation{Dept. of Engineering Physics, University of Wisconsin--Madison, USA}
\author{S. Gallian}
  \affiliation{Dept. of Electrical and Computer Engineering, University of Wisconsin--Madison, USA}
\author{E.~T. Hinson}
  \affiliation{Dept. of Engineering Physics, University of Wisconsin--Madison, USA}
\author{F.~A. Volpe}
  \affiliation{Dept. of Engineering Physics, University of Wisconsin--Madison, USA}

\date{\today}

\begin{abstract}
The ordinary--extraordinary (O--X) mode conversion is modeled with the aid of a 2D full-wave code in the {\sc Pegasus} Toroidal Experiment as a function of the launch angles.
It is shown how the shape of the plasma density profile in front of the antenna can significantly influence the mode conversion efficiency and, thus, the generation of electron Bernstein waves (EBW). It is therefore desirable to control the density profile in front of the antenna for successful operation of an EBW heating and current drive system.
On the other hand, the conversion efficiency is shown to be resilient to vertical displacements of the plasma as large as $\pm10$~cm.
\end{abstract}

\pacs{52.35.-g 52.35.Hr 52.35.Mw 52.50.Sw 52.55.Fa}

\maketitle

\section{Introduction}
In fusion experiments, it is a common method to heat the plasma by means of microwaves resonating with the electron cyclotron frequency $\omega_{ce}$ or its harmonics. The underlying mechanism is well understood~\cite{Erckmann_1994}, it allows to heat the electrons and drive significant currents. If, however, the plasma density exceeds the cutoff density for the frequency of the injected microwave, it is reflected before reaching the resonance and heating the plasma. This problem can be overcome by heating at higher harmonics of the cyclotron frequency, which requires electron temperatures of several keV to be efficient~\cite{Bornatici_1983}. Another approach is the utilization of electron Bernstein waves (EBWs). These are electrostatic waves that cannot propagate in vacuum and therefore need to be coupled to electromagnetic waves via mode conversion. No density cutoff exists for EBWs, and they are very well absorbed at $\omega_{ce}$ and its harmonics, even for low temperatures. Details about the physics and applications of EBWs can be found in a recent review article~\cite{Laqua_2007}.

EBW heating has been successfully demonstrated in stellarators~\cite{Laqua_1997} and tokamaks~\cite{Mueck_2007}. Especially in \emph{spherical} tokamaks, there is a need for starting and sustaining a plasma with reduced induction current since their geometry leaves only little space for a central transformer coil. Therefore, other current generation mechanisms have to be applied during the current ramp-up phase to save magnetic flux from the small solenoid. EBW current drive is one of such mechanisms, as demonstrated in TST-2~\cite{Shiraiwa_2006} and MAST~\cite{Shevchenko_2007}. EBW heating is also useful during the ramp-up phase, as it increases the electron temperature and decreases the collisionality, thus resulting in more effective induction. Helicity injection~\cite{Battaglia_2009,Redd_2009} and fast wave heating and current drive~\cite{Garstka_2006} are also expected to benefit from electron heating by EBWs in {\sc Pegasus}. Synergistic combinations of these start-up techniques have the potential to create a suitable plasma target to hand over to Ohmic heating and neutral beam injection in larger spherical tokamaks including a possible future component test facility~\cite{Peng_2005}.

In NSTX, the effect of collisions on the propagation of EBW has been studied experimentally~\cite{Diem_2009} and numerically~\cite{Preinhaelter_2006,Urban_2009}.
For the {\sc Pegasus} Toroidal Experiment, numerical studies of EBW propagation from the conversion layer and damping at the outermost Doppler-shifted electron cyclotron resonance (ECR) have been performed in the past, mostly by means of the GENRAY ray tracing code and the CQL3D Fokker-Planck code~\cite{Garstka_2006}. This paper investigates the {\it coupling} of externally launched electromagnetic waves with EBWs, using the 2D full-wave code IPF-FDMC~\cite{Koehn_2010phd}. The full-wave approach is indispensable to properly model the propagation of the O- and X-waves and their conversion into EBW because the vacuum wavelength of the injected microwaves, $\lambda_0\approx12.2$~cm, is comparable with the size of the plasma (average plasma minor radius $a\approx40$~cm).

The efficiency of the coupling is studied as function of the toroidal and poloidal injection angle. The role of the shape of the plasma density profile in front of the antenna is investigated and the possibility of heating at higher harmonic of the ECR frequency is explored. Thus, essential information for the design of an EBW heating system can be gained.

This paper is organized as follows: in section~\ref{s:pegasus}, the {\sc Pegasus} Toroidal Experiment is introduced and the equilibrium used in the modeling is described. The full-wave code IPF-FDMC is described in section~\ref{s:fullwave_intro}. Plasma and wave parameter scans in 1D and 2D simulations are presented and discussed in sections~\ref{s:fullwave_1D} and \ref{s:fullwave_2D}, respectively. We scan the density profile, the vertical position of the plasma, the injection angles, the polarization and the frequency of the injected microwave. Section~\ref{s:summary} summarizes and concludes this paper.

\section{The {\sc Pegasus} Toroidal Experiment}\label{s:pegasus}
\subsection{General properties}
The {\sc Pegasus} Toroidal Experiment~\cite{Garstka_2003} is a low aspect ratio facility with values of $A=R/a=1.15-1.3$, where $R$ and $a$ are the major radius and the average plasma minor radius, respectively. It was designed to explore the operational regime for spherical tokamaks with high toroidal beta
plasmas in the limit of $A\rightarrow 1$~\cite{Garstka_2006}. EBW heating can assist during the start-up phase of {\sc Pegasus} as it has been done in MAST~\cite{Shevchenko_2010}. The plasma core can be heated-up and EBW heating can provide a tool to control the current profile. Furthermore, as discussed in the introduction, synergies are possible between EBW heating and helicity injection~\cite{Battaglia_2009,Redd_2009} and between EBW heating and fast wave heating~\cite{Garstka_2006}.

\begin{figure}
  \includegraphics[width=.45\textwidth]{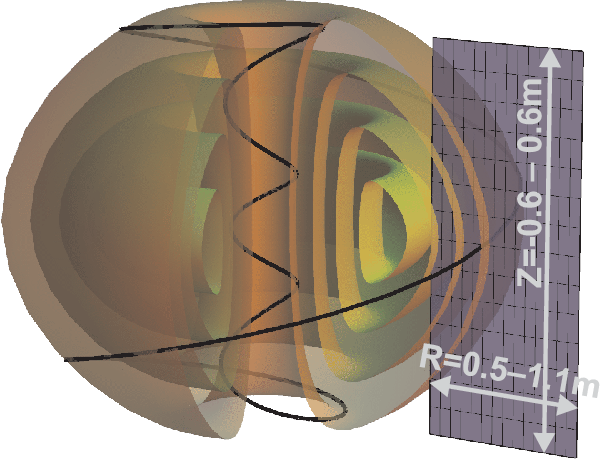}
  \caption{\label{fig:calculation_grid} {\sc Pegasus} flux surfaces, including a field line on the $q=5$ flux surfaces, with $q$ being the safety factor. The simulation plane is indicated on the right, the antenna is located in the equatorial plane at $R=1$~m.}
\end{figure}

EBW emission, heating and current drive experiments are under consideration for {\sc Pegasus}. An antenna would be installed for this purpose in one of the 12 equatorial ports. The diameter of the port (40 cm) puts a constraint on the maximum diameter of the injected or collected microwave beam. Figure~\ref{fig:calculation_grid} shows various {\sc Pegasus}' flux tubes and the orientation of the simulation plane.

Typical magnetic field strengths are on the order of 0.1~T in the core, making 2.45~GHz a good candidate for central EBW heating at the fundamental harmonic. Higher commercial and industrial standards, such as 3.6 and 5.5~GHz, are also of interest, for $2^{\mbox{nd}}$ harmonic EBW heating in the present configuration, or for $1^{\mbox{st}}$ harmonic heating in a possible {\sc Pegasus} upgrade to higher magnetic fields. Typical densities in the plasma center are on the order of $n_e=10^{19}$ m$^{-3}$ and electron temperatures up to $T_e\approx 300$~eV are achieved.

\subsection{Equilibrium used in the modeling}
\begin{figure}
  \includegraphics[width=.45\textwidth]{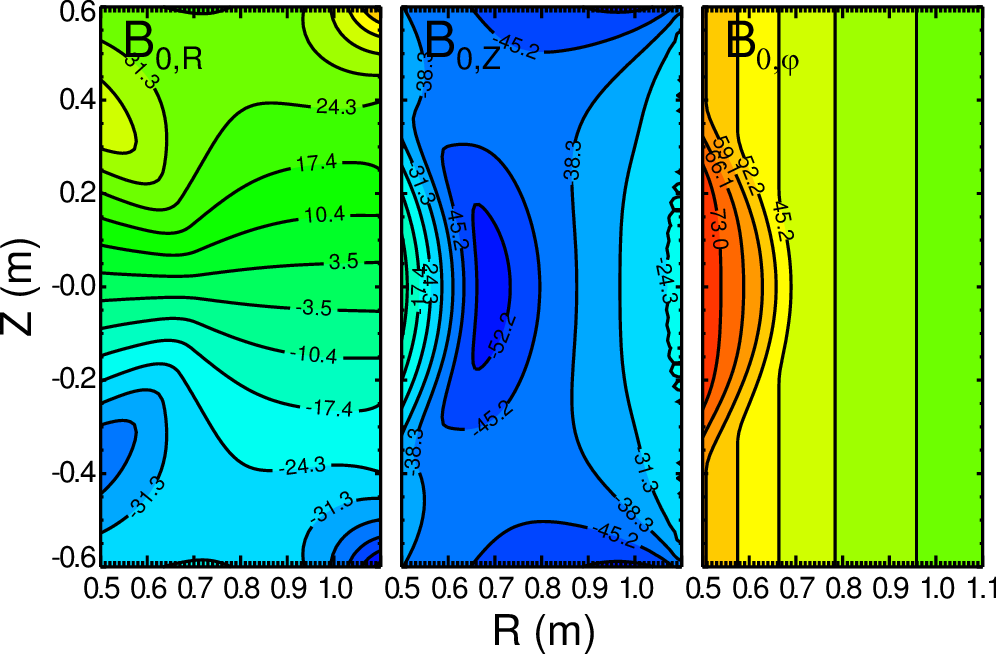}
  \caption{\label{fig:KFIT_B0_theta090} {\sc Pegasus} magnetic field adopted for the full-wave modeling from the KFIT equilibrium code. Shown are, respectively, the contours of the radial, vertical and toroidal component in a poloidal cross section. The values of the magnetic field strength are given in mT.}
\end{figure}

The KFIT equilibrium code~\cite{Sontag_2008} was used to model high performance plasma targets (of plasma current $I_P=150-300$~kA) for EBW experiments. Contour plots of the three components of the magnetic field in the poloidal plane as obtained from KFIT are shown in Fig.~\ref{fig:KFIT_B0_theta090}. Microwaves will be injected from the position $R=1$~m and $Z=0$~m.

\begin{figure}
  \includegraphics[width=.45\textwidth]{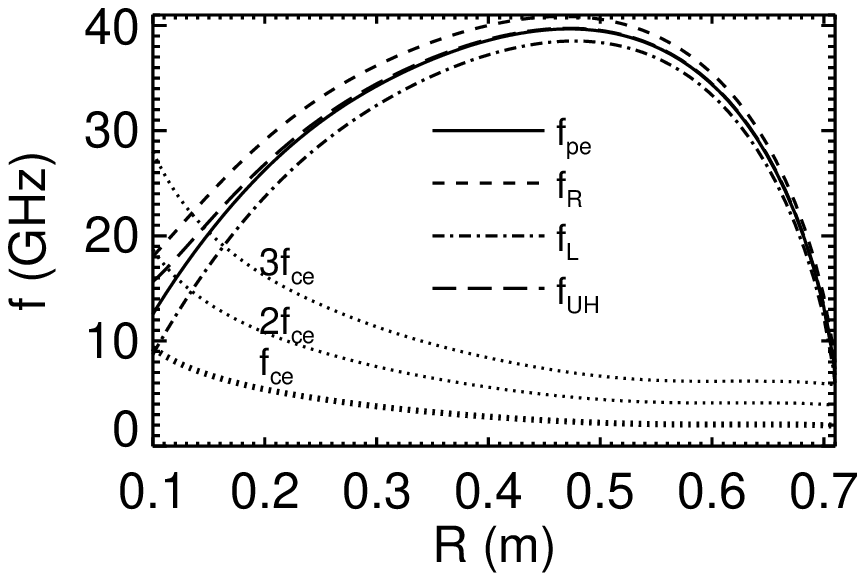}
  \caption{\label{fig:1D_cutRes} Cutoffs and resonances inside the LCFS as function of the radial coordinate for $Z=0$~m, different line styles represent different cutoffs and resonances, as labeled in the plot.}
\end{figure}

\begin{figure}
  \includegraphics[width=.9\textwidth]{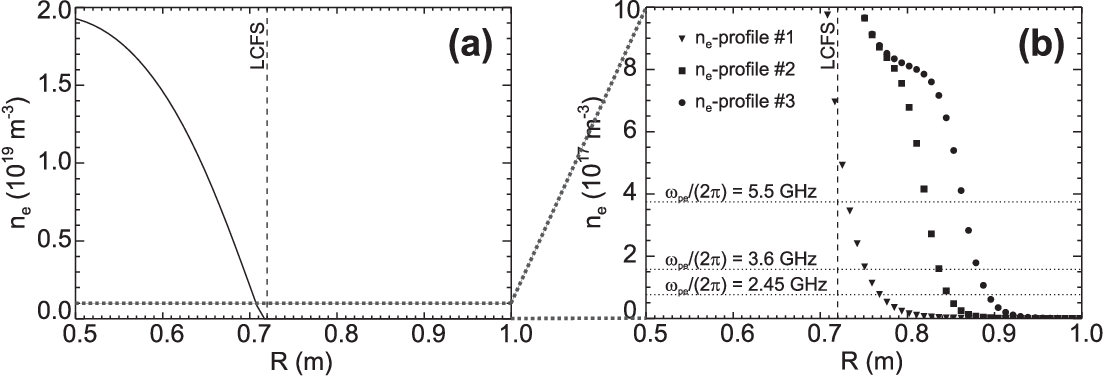}
  \caption{\label{fig:1D_density_profiles}Radial density profile at $Z=0$~m as used in the full-wave modeling: (a) assumed core profile and (b) edge profiles assumed on the basis of Langmuir probe measurements with the cutoff densities for different microwave frequencies marked by horizontal lines.}
\end{figure}

A reasonable plasma pressure profile \emph{inside} the last close flux surface (LCFS), which, for this equilibrium, is located at $R=0.72$~m for $Z=0$~m, serves as input constraint for the KFIT code. With the corresponding density profile and the values of the magnetic field strengths,
the characteristic cutoff and resonance frequencies can be calculated. They are shown in Fig.~\ref{fig:1D_cutRes} as function of the radial coordinate $R$. The value of the electron cyclotron frequency $f_{ce}$ is below the values of the cutoff frequencies in the complete cross section. Hence, the cyclotron frequency is shielded by cutoffs, and the entire {\sc Pegasus} plasma from the LCFS inwards, can be referred as \emph{overdense}.
It is also visible that the cutoff frequency at the LCFS is still well above 2.45~GHz. Thus, the O-mode cutoff layer (and potential O--X mode conversion layer) for 2.45~GHz lies well \emph{outside} the LCFS. In the mode conversion region, in the absence of density data from KFIT, assumptions about the shape of the density profile are necessary.
These assumptions are very delicate because
the shape of the profile, and in particular the density gradient length in the mode conversion region, play a crucial role in the mode conversion efficiency. Due to this sensitivity issue, different profiles were used in modeling the mode conversion region. Figure~\ref{fig:1D_density_profiles}a shows the assumed radial density profile at $Z=0$~m.
This is the internal density profile valid \emph{inside} the LCFS. It was extrapolated \emph{outside} the LCFS, up to the antenna, located at $R=1$~m.
Three different profiles, which are based on preliminary Langmuir probe measurements recently performed in the scrape-off layer, were used for the simulations presented in this paper and are shown in Fig.~\ref{fig:1D_density_profiles}b:
\begin{eqnarray}
  n_{e,1}(\rho) &=& n_{e,0}\exp\left\{- \left(\frac{\rho}{w_1}\right)^{\alpha} \right\} \label{eq:density1}\\
  n_{e,2}(\rho) &=& n_{e,1}(\rho) - b_1 \tanh \left\{ b_2 \left(\rho-\rho_2\right) \right\} + b_1 \label{eq:density2}\\
  n_{e,3}(\rho) &=& n_{e,1}(\rho) - b_1 \tanh \left\{ b_2 \left(\rho-\rho_3\right) \right\} + b_1,\label{eq:density3}
\end{eqnarray}
where $\rho$ is the normalized radius, $n_{e,0}=2\cdot 10^{19}$~m$^{-3}$, $w_1=0.7$, $\alpha=3.1$, $\rho_2=1.37$, $\rho_3=1.48$, $b_1=4\cdot 10^{17}$~m$^{-3}$ and $b_2=16$. In the experiment, it is speculated that such different shapes can be actively realized by different positions of a limiter placed closed to the antenna (i.~e. outside of the LCFS).

\section{The full-wave code IPF-FDMC}\label{s:fullwave_intro}
From Fig.~\ref{fig:1D_density_profiles}b, it can be deduced that, in {\sc Pegasus}, the density gradient length scale $L_n$, normalized to the vacuum wavelength, can become as small as $k_0L_n\approx2$ in the mode conversion region. For these steep profiles, with respect to the wavelength of the injected microwave, the geometric optics assumptions are not valid. It is therefore not possible to estimate the O--X--B mode conversion efficiency by analytic formulas~\cite{Preinhaelter_1973,Weitzner_1979,Mjolhus_1983} based on the validity of these assumptions. In a number of recent theoretical papers~\cite{Gospodchikov_2008,Weitzner_2004,Popov_2011} the O--X conversion has been investigated using a reduced system of wave equations in the vicinity of the conversion region coupled to a geometrical optics description in the rest of the plasma. Since in our case the validity of the geometrical optics assumptions breaks down as soon as the injected beam reaches the plasma, these models can also not be applied here. Hence, a full-wave description is necessary to model the mode conversion process. The code IPF-FDMC provides such a modeling.

In this section, the conversion process that results in the generation of an EBW is briefly described followed by a description of the full-wave code IPF-FDMC. Results of modeling the mode conversion process in {\sc Pegasus} are described in sections~\ref{s:fullwave_1D} and \ref{s:fullwave_2D}.

\subsection{O--X--B mode conversion scheme}
The O--X--B mode conversion scheme was conceived in 1973 by Preinhaelter and Kopeck$\acute{\mbox y}$~\cite{Preinhaelter_1973} as a method to heat overdense plasmas. In this scheme, an O-mode needs to be injected from the low field side at an optimum angle with respect to the background magnetic field. This O-mode is converted into an X-mode in the vicinity of the O-mode cutoff layer, which then propagates outwards until it reaches the upper-hybrid resonance (UHR) layer. There, it converts into an EBW, propagating backwards, which can then be absorbed at the ECR and its harmonics. Non-optimum injection angles lead to less efficient conversion, due to partial reflection from the cutoff layer or, for strongly non-optimum angles, refraction in the non-uniform plasma.


\subsection{The full-wave code IPF-FDMC}\label{s:fullwave_IPFFDMC}
The full-wave code IPF-FDMC
is a finite-difference time-domain (FDTD) code, solving Maxwell's equations, coupled with an equation for the current density $\mathbf J$, which is obtained from the fluid equation of motion of the electrons, on a Cartesian grid. Details on the FDTD method in general can be found in Ref.~\cite{TafloveHagness_2005}. The system of equations to be solved for each time step reads:
\begin{eqnarray}
  \frac{\partial}{\partial t} \mathbf B &=& -\nabla \times \mathbf E\label{eq:ipffdmc_B}\\
  \frac{\partial}{\partial t} \mathbf E &=& c^2\nabla \times \mathbf B - \frac{1}{\epsilon_0}\mathbf J\label{eq:ipffdmc_E}\\
  \frac{\partial}{\partial t} \mathbf J &=& \epsilon_0\omega_{pe}^2\mathbf E - \omega_{ce} \mathbf J \times \hat{\mathbf B}_0 - \nu \mathbf J\label{eq:ipffdmc_J},
\end{eqnarray}
where $\mathbf E$ and $\mathbf B$ are respectively the electric and magnetic field of the wave, $\omega_{pe}$ is the electron plasma frequency, $\omega_{ce}$ the electron cyclotron frequency and $\nu$ a collisional frequency.
The $\nu \mathbf J$ term is responsible for collisional damping of the wave. While this damping mechanism is real and non-negligible, here it is artificially enhanced to prematurely cause the complete damping of the wave before its wavelength becomes too short (when the X-wave is approaching the UHR) for the adopted numerical grid to resolve it. Note that this enhancement does not change the conversion efficiency~\cite{Koehn_2008}. A more realistic picture is that the slow X-wave experience some collisional damping at the UHR and converts into an EBW that propagates towards the plasma core and is cyclotron-damped in the vicinity of the outermost ECR. This treatment, however, would require a finer numerical grid, finite Larmor radius corrections and a model for the cyclotron-damping of EBWs suitable for full-wave calculations or, alternatively, the coupling of the O--X full-wave solution with a ray tracing code for the propagation and damping of EBWs~\cite{Volpe_2003}. Such improvements are left for future work.

Here, $\mathbf B$, $\mathbf E$ and $\mathbf J$ are three dimensional vector fields and all three components in the $R$, $Z$ and $\varphi$ direction are calculated and advanced in time, although they are treated as functions of $R$ and $Z$ only. In other words, the plasma inhomogeneity is 2-dimensional and the problem is invariant under translation in the transverse direction. The code was successfully used to model the O--X--B mode conversion process in the TJ-II stellarator~\cite{Koehn_2008} and the RFX-mod reversed field pinch~\cite{Bilato_2009}. Furthermore, microwave heating (although not by EBW) of the TJ-K stellarator~\cite{Koehn_2010} has been modeled with IPF-FDMC and a predecessor of the code has been applied to the WEGA stellarator~\cite{Podoba_2007}. An antenna is simulated in the code by adding a time-harmonic field to a certain position on the numerical grid.
More details about this mechanism and about the code in general can be found in Ref.~\cite{Koehn_2010phd}.

Equations~(\ref{eq:ipffdmc_B})--(\ref{eq:ipffdmc_J}) include only cold plasma effects, which are sufficient to model the O--X mode conversion. In the vicinity of the UHR, the X-mode becomes more and more electrostatic, its wavelength becomes shorter and the cold plasma formulation breaks down. To resolve this singularity, it is in principle possible~\cite{Koehn_2008} to include first order finite Larmor radius corrections~\cite{Ram_1996} in the code to account for the X--B conversion. However, here, electron collisions are used to damp the X-mode around the UHR, since the inclusion of the X--B conversion would significantly increase the computational time. The reason is that the wavelength of the EBW is comparable with the electron Larmor radius, i.~e. much smaller than the vacuum wavelength, thus requiring a much finer numerical grid.
Therefore, we restrict ourselves to the O--X conversion in this paper.
Note that some effects that deteriorate the overall conversion efficiency are included in the present model, but some others are not. A deteriorating mechanism is the coupling between internal slow X-waves (excited via the O--X conversion) into external fast X-waves, leaving the plasma~\cite{Volpe_2003phd}. Slow and fast X-waves are normally separated by an evanescent layer which, however, can be very thin in the low magnetic field edge of a spherical tokamak. This effect is automatically taken into account by the full-wave code. An effect, not taken into account by the code, is the parametric decay in waves of different, generally undesired frequencies~\cite{Laqua_2007}. The effect of density fluctuations on the O--X conversion, which are expected to locally and temporarily modify the optimal direction for efficient mode conversion, is also not taken into account, but will be the subject of a future study. On average, this effect is a reduction of the conversion efficiency.
Hence, the O--X conversion efficiency deduced from the present simulations has to be considered as an upper limit for the actual overall O--X--B conversion efficiency.
Note also, however, that the X--B conversion does not introduce any additional angular dependence. Therefore, the optimal direction and angular tolerance of the actual O--X--B process coincide with the O--X optimal direction and angular tolerance obtained from the present simulations.

The vacuum wavelength of the injected microwave is set to be 256 grid points on the numerical grid and the normalized collision frequency lies in the range $10^{-5}\le \nu/\omega_0 \le 10^{-3}$ (see Ref.~\cite{Koehn_2008} for details).

\section{1D simulations}\label{s:fullwave_1D}
Calculations on a 1D grid, which require significantly less computational resources than calculations on a 2D grid,
were performed first. The plasma density is taken to be a function of the radial coordinate $R$, only, and the magnetic field is taken to be perpendicular to the radial coordinate with a constant value of $\omega_{ce}/\omega_0=0.6$.
The injected microwaves correspond to plane waves in this case. Although the experimental situation is not described properly by these simplifications, the results from these calculations can serve to set constraints on parameters for the 2D calculations.

\begin{figure}
  \includegraphics[width=.45\textwidth]{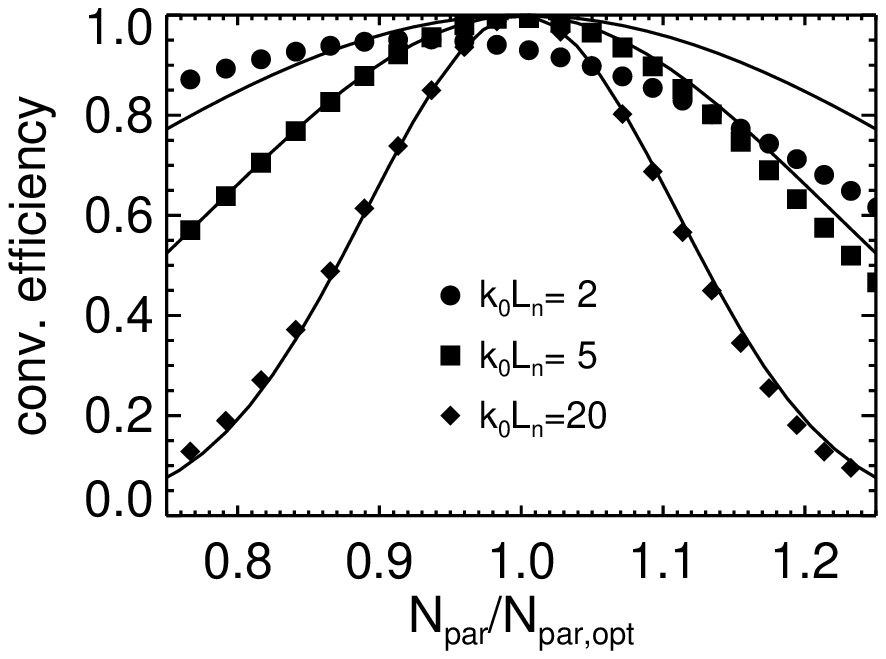}
  \caption{\label{fig:fullwave_1D_eta_Nz} Conversion efficiency as a function of the parallel component of the normalized refractive index, where \emph{parallel} refers to the direction of the magnetic field. The different symbols correspond to results obtained from full-wave simulations for different values of the density length scale normalized to the vacuum wavelength, $k_0L_n$, as labeled in the plot. The solid lines indicate the corresponding solution from the Mj\o lhus equation\cite{Mjolhus_1983}.}
\end{figure}

For small values of $k_0L_n$, Mj\o lhus' formula for the conversion efficiency as a function of the injection angle~\cite{Mjolhus_1983} is no longer valid. This is nicely illustrated in Fig.~\ref{fig:fullwave_1D_eta_Nz}, where simulations for three different values of $k_0L_n$ are compared with the corresponding solution from Mj\o lhus' equation. For these 1D simulations, the density profiles were taken to be of parabolic shape. In the case considered here, the variation between the analytical solution in the WKB limit and the full-wave solution is not very strong for $k_0L_n\ge5$. However, at small density gradient length, for example such that $k_0L_n=2$, the WKB limit and Mj\o lhus' formula are no longer valid. This results in the discrepancy between the values obtained from the simulations and Mj\o lhus' formula that can be seen in Fig.~\ref{fig:fullwave_1D_eta_Nz}.
The general trend that for steeper profiles the conversion efficiency becomes less sensitive to an angular mismatch can also be clearly seen.

\begin{figure}
  \includegraphics[width=.45\textwidth]{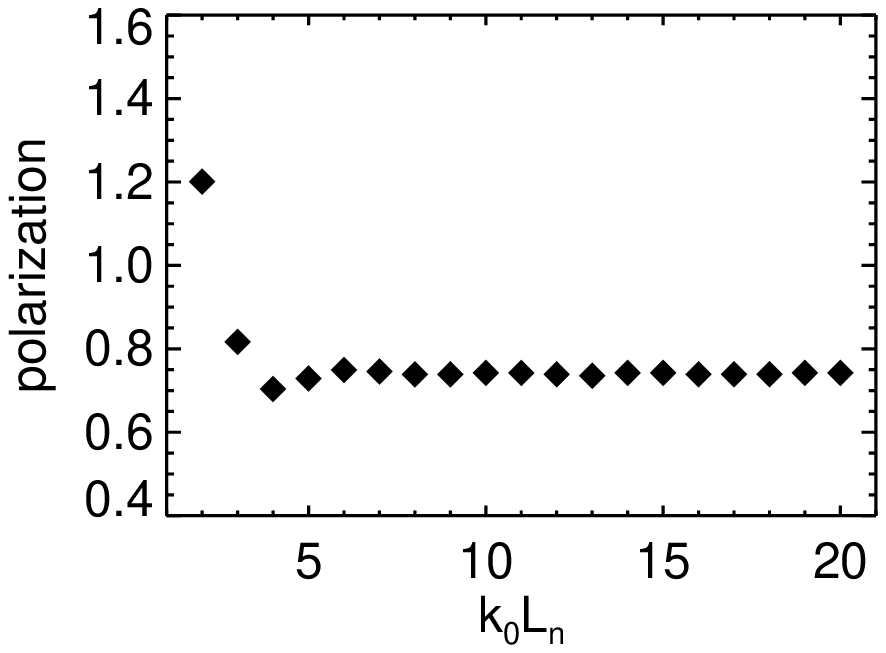}
  \caption{\label{fig:fullwave_1D_pola_k0Ln} Polarization of injected microwave beam, $E_{R}/E_{\perp}$ with $E_{R}$ the wave electric fields along the calculation grid and $E_{\perp}$ the field perpendicular to $R$ and to the direction of the magnetic field, for optimum conversion efficiency, obtained from full-wave simulations, as function of the normalized density gradient length.}
\end{figure}

If the density profile becomes too steep, the polarization of the injected microwave needs to be re-adjusted to obtain maximum conversion efficiency~\cite{Igami_2006}. This effect is illustrated in Fig.~\ref{fig:fullwave_1D_pola_k0Ln}, where the optimum polarization $E_{R}/E_{\perp}$ (with $E_{R}$ the wave electric field along the calculation grid and $E_{\perp}$ the field perpendicular to $R$ and to the direction of the magnetic field) is plotted as function of $k_0L_n$. The values were obtained from a series of simulations in which the polarization was scanned in order to find its optimum value. As one can see, for values of $k_0L_n\le5$, the polarization differs from its asymptotic value, which can be calculated analytically~\cite{Hansen_1985} for this configuration to $E_{R}/E_{\perp}\approx0.75$.

\begin{figure}
  \includegraphics[width=.45\textwidth]{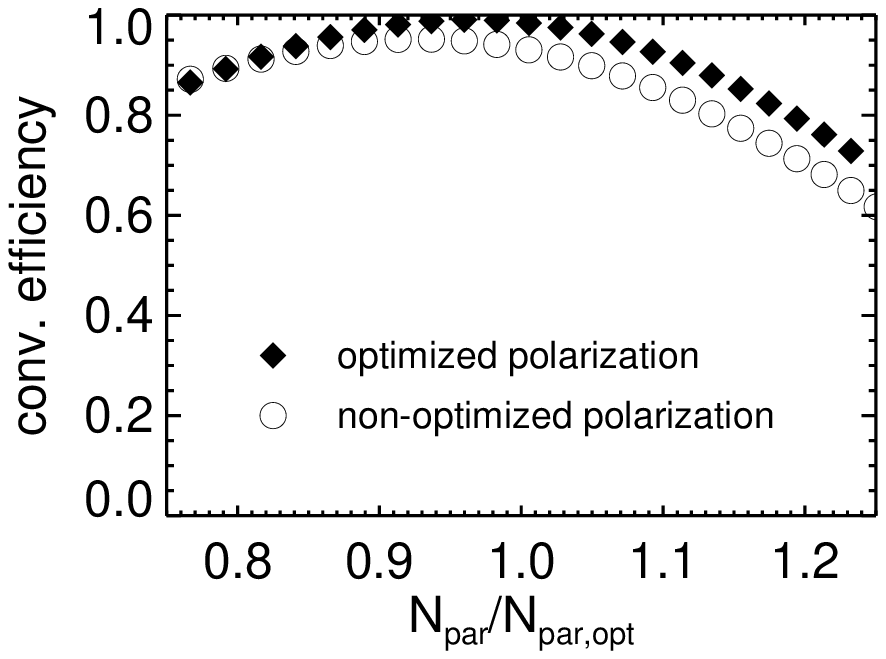}
  \caption{\label{fig:fullwave_1D_polaopt_vs_polanonopt} Conversion efficiency as function of the parallel component of the normalized refractive index obtained from simulations with $k_0L_n=2$ with optimized and non-optimized polarization of the injected beam.}
\end{figure}

To check the \emph{relevance} of the aforementioned polarization adjustment, two simulations were performed for steep profiles with $k_0L_n=2$, one with the adjusted polarization and the other with the non-adjusted polarization. Their results are shown in Fig.~\ref{fig:fullwave_1D_polaopt_vs_polanonopt}.
One can clearly see that optimizing the polarization has only a fairly small effect on the conversion efficiency and on the optimal injection angle, or, in other words, the conversion efficiency is relatively stable against polarization mismatch.

\section{2D simulations}\label{s:fullwave_2D}
From the 1D model discussed in the previous section, it could be deduced that for steep density profiles, the angular window to achieve maximum mode conversion efficiency is fairly large. That result is confirmed here in 2D, taking into account
the geometry of the plasma and the finite size of the beam.

The flux surfaces in {\sc Pegasus} are significantly curved, as can be seen in Fig.~\ref{fig:calculation_grid}. Their curvature radius in the region of interest is on the order of 50~cm and non-negligible on the transverse length scale of a realistic 2.45~GHz beam. A previous work~\cite{Koehn_2008} showed that in the presence of curved flux surfaces it is important for wavefronts to match their curvature to the curvature of the conversion layer.
Hence, a Gaussian microwave beam emitted from the antenna with the beam waist located inside the plasma is considered in the simulations (for details on Gaussian beam properties, see e.~g. Ref.~\cite{Siegmann_1971}). Following a previous, internal design study, the emitting microwave antenna is located at $R=1$~m with the beam waist located at $R=0.7725$~m and a beam radius (at the waist) of $w_0=0.67\lambda_0$.

\begin{figure}
  \includegraphics[width=.8\textwidth]{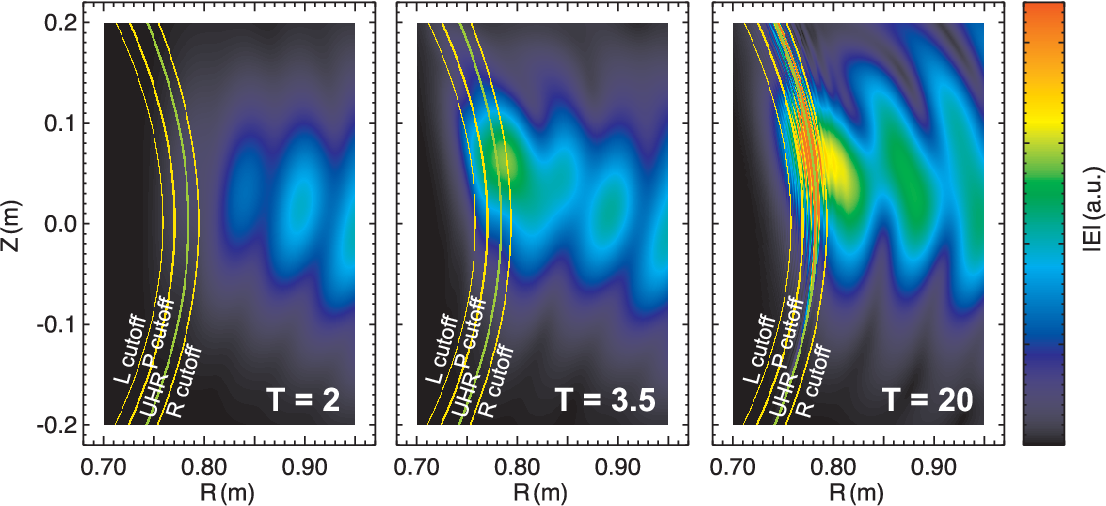}
  \caption{\label{fig:fullwave_2D_snapshots} Snapshots of the absolute value of the wave electric field, $|E|$, with the time T given in units of oscillation periods in the lower right corner of each plot.}
\end{figure}

In the 2D simulations, the poloidal and toroidal injection angle can be scanned in order to find the maximum conversion efficiency. Figure~\ref{fig:fullwave_2D_snapshots} shows three snapshots at different times of the absolute value of the wave electric field for a poloidal and toroidal injection angle of 10\textdegree\ and 0\textdegree, respectively. Density profile \#1 (see Fig.~\ref{fig:1D_density_profiles}b and Eq.~(\ref{eq:density1})) was used in this simulation. The positions of the characteristic frequency layers for 2.45~GHz are given in the plot. In the first snapshot, taken after $T=2$ oscillation periods, the wave has not yet reached the conversion layer. Its focused phase fronts can be clearly seen. At $T=3.5$, the wave has reached the conversion layer and one can see a complicated interference pattern building up between the incoming and the reflected (i.~e. not converted) wave. After $T=20$ oscillation periods, a steady state situation has been reached and a conversion efficiency of 33\% is found. The generated X-mode, which is visible by its enhanced wave electric field and the small scale structure, is damped at the UHR, since no X--B conversion is included in this simulation (see Sec.~\ref{s:fullwave_IPFFDMC}).

\begin{figure}
  \includegraphics[width=.8\textwidth]{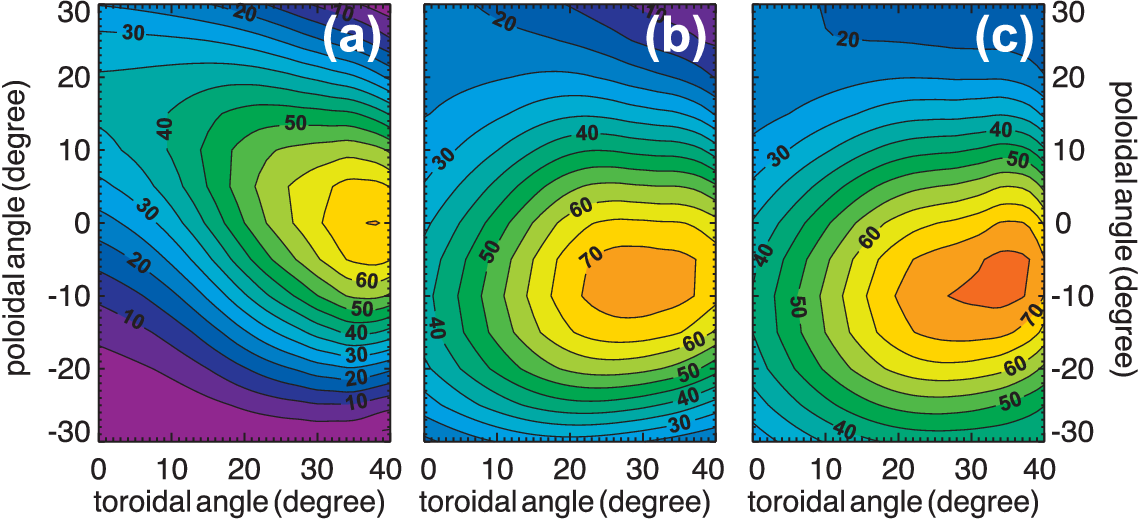}
  \caption{\label{fig:fullwave_2D_OXcontour_ne267} O--X conversion efficiency as a function of the poloidal and toroidal injection angle for (a) density profile \#1 (b) \#2 and (c) \#3 (see profiles in Fig.~\ref{fig:1D_density_profiles}).}
\end{figure}

In order to check for the influence of the shape of the density profile on the conversion efficiency, for each of the three profiles shown in Fig.~\ref{fig:1D_density_profiles}b, more than 500 runs have been performed, each with a different combination of poloidal and toroidal injection angle. After the steady state situation has been reached, the conversion efficiency can be deduced. It is plotted in Fig.~\ref{fig:fullwave_2D_OXcontour_ne267} as function of the toroidal and poloidal injection angle.
The maximum conversion efficiency is found at different injection angles for different density profiles. This is especially true for profile \#1, as opposed to the two other profiles. The general shape of the angular window differs also when comparing the results from profile \#1 with the two others: the window becomes broader with the density profile moving closer to the antenna. The highest conversion efficiencies of 75\% are also achieved for profile \#3, in which the distance between the mode conversion layer and the antenna is the shortest. Density profile \#2 and \#3 have shorter gradient lengths in the mode conversion region, and, thus, smaller values of $k_0L_n$, which results in the observed broader angular window (c.~f. Fig.~\ref{fig:fullwave_1D_eta_Nz}).

These results clearly illustrates the importance of the knowledge of the actual density profile in front of the antenna for efficient EBW coupling. It was also shown how smaller values of $k_0L_n$ result in larger angular windows for efficient mode conversion (the same result was found in the 1D simulations). Hence, steeper profiles in front of the antenna seem to be preferable in this case, although even steeper profiles would lead to a deterioration of the conversion efficiency, as described in Sec.~\ref{s:fullwave_IPFFDMC}.

\subsection{Stability against vertical displacement of the plasma}
Highly elongated plasmas like {\sc Pegasus} can be subject to vertical displacements during the discharge. It is therefore of interest, how an EBW heating system couples with a plasma that moves vertically.
Thus, simulations have been performed, where the antenna was shifted 10~cm upwards and 10~cm downwards relative to the plasma, corresponding to shifting the plasma 10~cm downwards and upwards, respectively.

\begin{figure}
  \includegraphics[width=.7\textwidth]{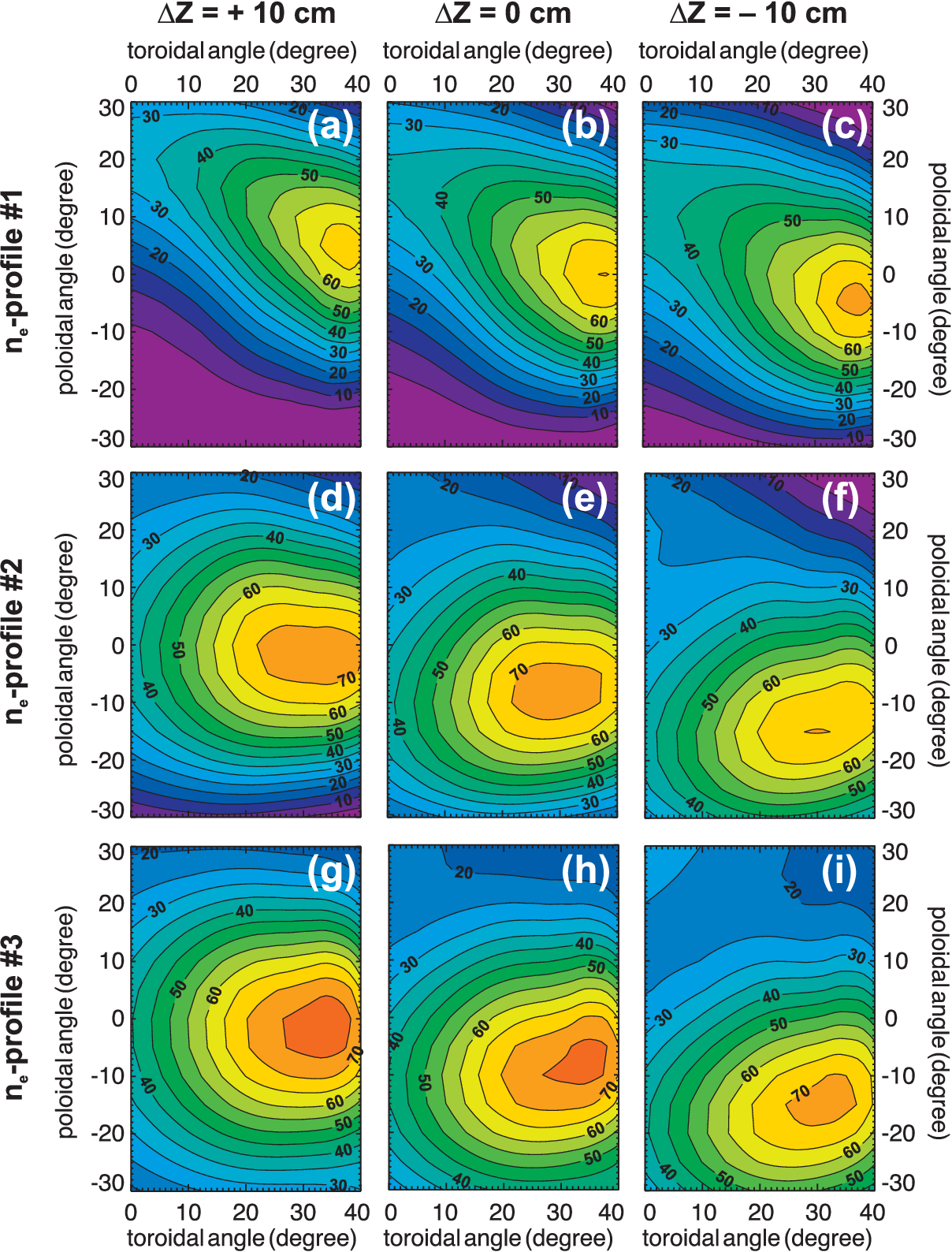}
  \caption{\label{fig:fullwave_2D_OXcontour_ne276vertDisplace} O--X conversion efficiency as a function of the poloidal and toroidal injection angle for (a)--(c) density profile \#1 with the plasma shifted 10~cm upwards, not shifted and shifted 10~cm downwards, respectively, (d)--(f) the same for density profile \#2 and (g)--(i) for density profile \#3.}
\end{figure}

Figure~\ref{fig:fullwave_2D_OXcontour_ne276vertDisplace} shows the O--X conversion efficiency for the three density profiles with a displacement of $+10$~cm, without and with $-10$~cm displacement. Note that such vertical displacements are extreme and rarely observed. Even under these extreme circumstances, the optimal launch angle in the vertical direction only varies by approximately 5\textdegree\ for the density profile \#1,  i.~e. by much less than the angular window width. For the two other profiles, the variation is slightly stronger, but still below 10\textdegree.
Hence, the plasma is relatively stable against vertical displacement and a reasonable amount of the injected microwave power would still be converted into an EBW when a small vertical displacement occurs.

\subsection{Mode conversion efficiency at higher microwave frequencies: 3.6 and 5.5~GHz}

\begin{figure}
  \includegraphics[width=.7\textwidth]{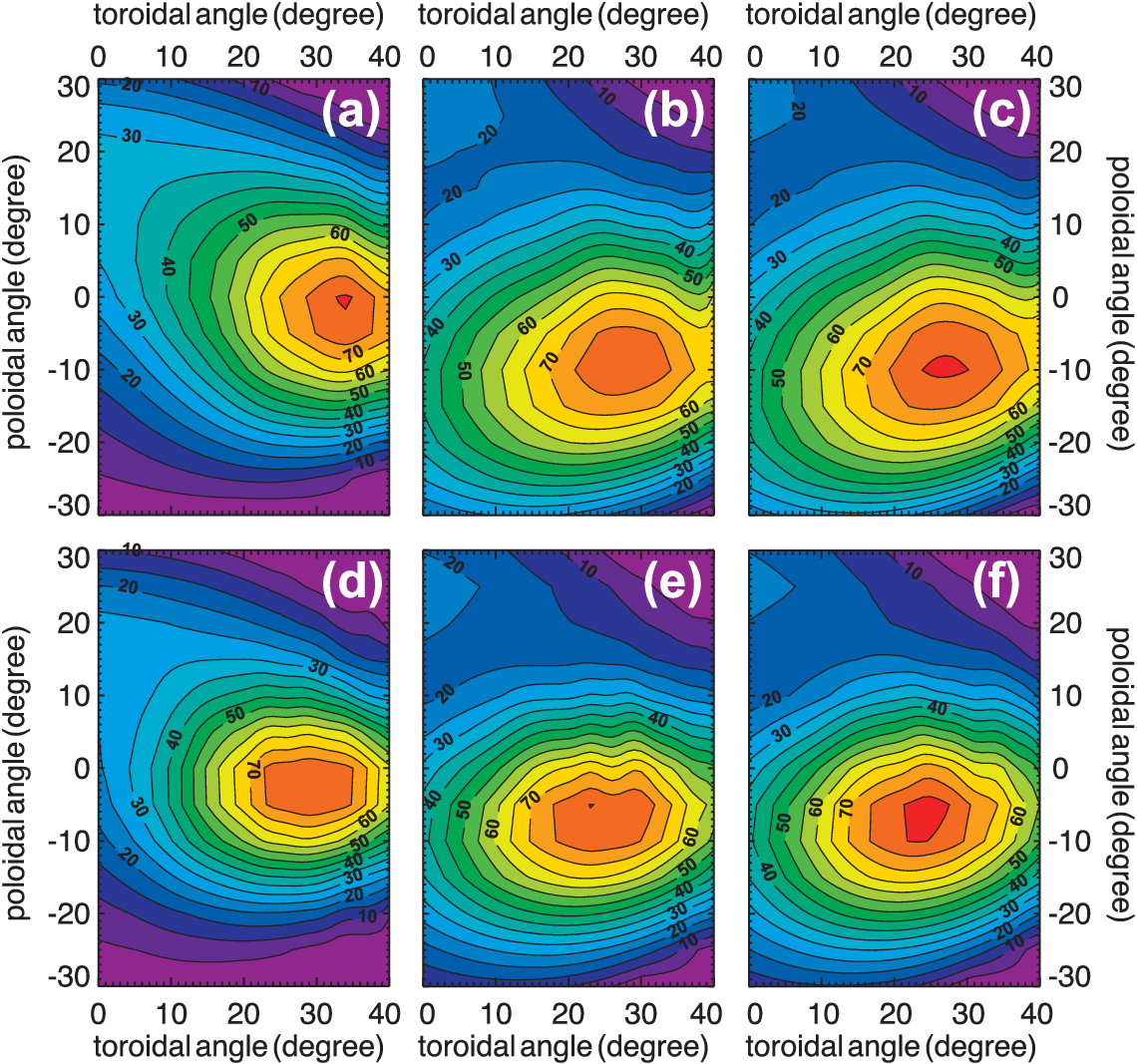}
  \caption{\label{fig:fullwave_2D_OXcontour_grid5D6D_ne276} O--X conversion efficiency as function of poloidal and toroidal injection angle for (a)--(c) a microwave frequency of 3.6~GHz and the density profiles \#1--\#3 and (d)--(f) a microwave frequency of 5.5~GHz and the same profiles.}
\end{figure}

One feature of EBWs is that they are very well absorbed at harmonics of the ECR frequency, in contrast to conventional ECR heating, which requires electron temperatures of several keV to be efficient for higher harmonic heating~\cite{Bornatici_1983}. For this reason, simulations with a microwave frequency of 3.6~GHz have been performed which corresponds to $2^{\mbox{nd}}$ harmonic heating. In Fig.~\ref{fig:fullwave_2D_OXcontour_grid5D6D_ne276}a--c, the obtained conversion efficiencies for the density profiles \#1--\#3 are shown, respectively. When comparing them with Fig.~\ref{fig:fullwave_2D_OXcontour_ne267}a--c, where the conversion efficiencies for 2.45~GHz are shown, one can see that slightly higher values (75--80\%) are achieved at the optimum injection angles, which, for their part, have barely changed. The angular window for high efficiencies has become smaller, which is due to the fact that with the increased value of $k_0\approx75.5$~m$^{-1}$ ($k_0\approx51.3$~m$^{-1}$ for 2.45 GHz), the value of $k_0L_n$ at the conversion layer has also increased, which results in a stronger sensitivity to the injection angle, as discussed in Sec.~\ref{s:fullwave_1D}.

A possible upgrade of {\sc Pegasus} to higher magnetic fields would require higher microwave frequencies, due to the increased ECR frequency. To check for the potential of such an upgrade on EBW heating, additional simulations were performed for 5.5~GHz. The conversion efficiencies obtained are shown in Fig.~\ref{fig:fullwave_2D_OXcontour_grid5D6D_ne276}d--f. One can see that the angular window has become slightly smaller due to the increased value of $k_0L_n$. The difference between the shape of the contours of the conversion efficiency for the three different density profiles is smaller as compared with the lower microwave frequencies. This is due to the reduced relative difference of $k_0L_n$ between the different profiles. The maximum efficiencies achieved for 5.5~GHz are similar to the case for 3.6~GHz. Thus, this scenario seems also to be suitable for EBW heating.

\section{Summary}\label{s:summary}
The O--X mode conversion process has been modeled with the full-wave code IPF-FDMC in the {\sc Pegasus} Toroidal Experiment. A frequency of 2.45~GHz has been chosen, since experiments are under consideration at this frequency. Different density profiles, based on preliminarily Langmuir probe measurements, were included in the simulations. All of these profiles are rather steep, with values of the normalized density gradient length of $k_0L_n\approx 2$ in the mode conversion layer. Maximum conversion efficiencies on the order of 70--75\% are found. The angular window for conversion efficiencies above 50\% is fairly large for all density profiles. However, they are of different shape, which illustrates that the knowledge of the shape of the actual density profile in the mode conversion region is important, even for such small values of $k_0L_n$, if an EBW heating and current drive system with a power level in the MW regime is considered.

An extreme vertical displacement of the plasma by $\pm 10$~cm, corresponding to an extreme instability, still results in high conversion efficiencies on the order of 65\%, if the injection angles are not adjusted. Hence, the O--X conversion is relatively stable against such displacements, thanks to the large angular windows mentioned above.

For $2^{\mbox{nd}}$ harmonic heating with 3.6~GHz, higher conversion efficiencies of 75--80\% are obtained, but the width of the angular window is slightly smaller, due to the increased value of $k_0L_n$ at the mode conversion layer. A potential upgrade of {\sc Pegasus} to higher magnetic field strengths would require higher microwave frequencies, such as 5.5~GHz, for which simulations yield similar conversion efficiencies as for the case with 3.6~GHz, but with yet smaller angular windows.

To conclude, the full-wave simulations showed that high O--X conversion efficiencies of up to 80\% can be achieved 
at {\sc Pegasus}. These estimates represents an upper limit for the overall O--X--B conversion efficiency, the reason being that
some degradation effects like the excitation of parametric instabilities during the X--B conversion are not included in the simulations yet. It should also be pointed out, however, that the X--B conversion does not introduce any further dependence on the launch angles. Therefore, the optimum angles for O--X conversion calculated in this work are also the optimum angles for EBW heating and current drive at {\sc Pegasus} by means of the O--X--B conversion.

\begin{acknowledgments}
Authors A.~K\"ohn and J.~Jacquot wish to express their gratitude towards the Erasmus Mundus program for financial support during their visit to the University of Wisconsin--Madison.
\end{acknowledgments}

\nocite{*}
\providecommand{\noopsort}[1]{}\providecommand{\singleletter}[1]{#1}%

\end{document}